\documentclass[amsmath,amssymb,eqsecnum,showpacs,nofootinbib,twocolumn]{revtex4-1}

\usepackage{chngcntr}

\usepackage{graphicx}
\usepackage{bm}
\usepackage{dcolumn}
\usepackage{color}

\newcommand{\rc}[1]{}

\newcommand{\be}{\begin{equation}}
\newcommand{\ee}{\end{equation}}
\newcommand{\bea}{\begin{eqnarray}}
\newcommand{\eea}{\end{eqnarray}}

\begin{document}

\counterwithout{equation}{section}

\author{Alessandro Fabbri}
\affiliation{Departamento de F\'isica Te\'orica and IFIC, Centro Mixto Universidad de Valencia-CSIC, C. Dr. Moliner 50, 46100 Burjassot, Spain.}
\email{afabbri@ific.uv.es}

\author{Roberto Balbinot}
\affiliation{Dipartimento di Fisica dell'Universit\`a di Bologna and INFN sezione di Bologna, Via Irnerio 46, 40126 Bologna, Italy,}
\email{balbinot@bo.infn.it}

\title{Ramp-up of Hawking radiation in Bose-Einstein condensate analogue black holes}

\begin{abstract}
Inspired by a recent experiment by Steinhauer and co-workers, we present a simple model which describes the formation of an acoustic black hole in a Bose-Einstein condensate,  allowing an analytical computation of the evolution in time of the corresponding density-density correlator. We show the emergence of analog Hawking radiation out of a ``quantum atmosphere'' region significantly displaced from the horizon. This is quantitatively studied both at $T=0$ and even in the presence of an initial temperature T, as is always the case experimentally.
\end{abstract}
\date{\today}
\maketitle



Hawking black hole (BH) evaporation \cite{hawking} can be understood as a pairs creation process in which a member of the pair is created outside the horizon constituting the thermal radiation emitted by the BH.
The other member, called partner, is created inside the horizon and has negative (Killing) energy. The members of the pair are entangled and this leads to correlations. The presence of a causal horizon in a BH prevents a direct measurement of these correlations.
This kind of particle-partner creation mechanism is however not peculiar to BHs \cite{unruh}.  
The same process is at work in various others physical settings (called analogue BHs), in particular in flows that from subsonic turn to supersonic (for a review, see \cite{ barcelo}). Sonic BHs constructed from Bose-Einstein condensates (BECs) are the most studied examples  \cite{gacz}.

In this case it has been predicted that the particle-partner correlations of Hawking radiation will be manifested by the presence of a stationary peak which appears at late time after the formation of the sonic horizon in the equal time density-density correlator when one point is taken outside the horizon and the other inside \cite{bffrc, cfrbf}. This striking feature has indeed be experimentally observed by Steinhauer  \cite{jeff2016, jeff2019} and this is the most stringent evidence of Hawking like (phonons in this case) radiation in a BEC.
Having an (almost) complete comprehension of stationary analogue BHs, research is now moving forward trying to understand what happens in the dynamical formation of time dependent horizons \cite{jeff-nuovo, silkeetal}. This is a new era in analogue gravity where one is  not just satisfied to find the stationary Hawking radiation, but one wants to understand where and when this radiation emerges.
For gravitational BHs it has been argued by various authors starting from Unruh \cite{unruh1977, schun, qatm1, qatm2, qatm3} that Hawking particles emerge from a region significantly  displaced from the horizon which has been named ``Quantum Atmosphere" by Giddings \cite{qatm1}. We will see that our analysis will give a strong support to it.

Our work is inspired by a recent work of Steinhauer and co-workers who have reported on an experiment which was able to follow through correlations measurements the time evolution of a BEC BH \cite{jeff-nuovo}. Starting from the formation of the sonic horizon they observed the ramp-up of Hawking radiation to its stationary regime
(see the first two phases in their Fig. 2).
The purpose of this paper is to show how, with the methods of Quantum Field Theory (QFT) in curved space, one can describe quite nicely the emergence of Hawking radiation in a BEC  by following the time evolution of the relevant density correlator toward its stationary configuration\footnote{A preliminary study in this direction can be found in Appendix C of \cite{mcp}.}. This is done by using a simple toy model of sonic hole formation which captures the relevant features of the ramp-up and has the advantage of allowing an analytical treatment which can also be extended to the case of a non vanishing ambient temperature of the BEC.

In the spirit of the gravitational analogy \cite{livrevrel}, phonons in a spatially one dimensional BEC can by described by a 3+1 dimensional massless scalar field $\delta \hat \theta$, representing the phase fluctuation of the condensate,  propagating in an acoustic metric associated to the one dimensional (along the $x$ axis) BEC flow:
\be ds^2 = \frac{n}{mc} [ -( c^2 - V^2 ) dt^2  - 2Vdtdx + dx^2 + dy^2 +dz^2 ]\ , \label{uno} \ee                                         
where $V(x)$ is the velocity of the flow and $c(x)$ the speed of sound,  $n(x)$ is the condensate density and $m$ the mass of a single atom.
The transverse size $l_\perp$ of the condensate is assumed to be much smaller than the healing length $\xi=\frac{\hbar}{mc}$, as it happens in the
experimental realization of Steinhauer. This allows to treat the system as effectively 1D.

The field $\delta \hat \theta$ satisfies
\be \Box \delta \hat \theta = 0 \ , \label{due}  \ee
where the covariant D'Alembert operator is constructed from the metric (\ref{uno}).
We will be interested in the correlator of the 1D density fluctuations $\delta \hat n^{(1)}$ which is constructed from $\delta \hat \theta$ as
\be \delta \hat n^{(1)}  =- \frac{n^{(1)}}{mc^2}(\partial_t +V\partial_x)  \delta \hat \theta\ , \label{tre}  \ee
where $n^{(1)}=nl_\perp^2$ is the 1D density of the condensate.
Relation (\ref{tre}) holds in the so called hydrodynamical approximation which is valid on longitudinal scales much bigger than the healing length  
$\xi$. 
This approximation is the core of the gravitational analogy in BECs. 

We shall consider the flow directed from right to left at a constant velocity $V$ ($<0$) and whose density $n$ is also constant. The profile for the speed of sound is assumed to be the following :
\bea \label{quattro}
\begin{cases}
c &= c_{in} , \ \ \   t<0   \\              
c &= |V|\left( 1+\frac{2}{3}\tanh \frac{3\kappa x}{2|V|}\right)            ,\ \ \    t>0\\
\end{cases}
\eea
where $c_{in}$ ($>|V|$) is a constant and $\kappa$ is also a constant. 
One can vary the speed of sound $c \ (=\sqrt{\frac{gn}{m}})$ for example by varying the  atom-atom interaction coupling $g$
using Feshbach resonances \cite{pi-st}, or as described in Ref. \cite{jeff2019}.
We can generalize the model by letting both c and V vary as it is the case in the set-up of the experiment described in \cite{jeff-nuovo}. \footnote{The physical characteristics of a condensate like the speed of sound are  routinely modulated in the laboratories, even suddenly, see for  example the detection of phonons in a BEC by the dynamical Casimir  effect performed by Westbrook and co-workers \cite{chris2012}.}
We believe however that our simple toy model is sufficient to reproduce at least qualitatively the relevant features recently observed by Steinhauer and co-workers.

Our choice describes a uniform subsonic flowing condensate for $t<0$. Instantaneously at t=0 a sonic BH forms: the flow remains subsonic for $x>0$, while it becomes supersonic for $x<0$. The horizon is at $x=0$ and $\kappa$ is its surface gravity.

For the reasons given before we shall be interested in the equal time density-density correlator
$ G_2^{(1)}(t;x,x')= \lim_{t\to t'} \langle \delta \hat n^{(1)}(t,x) \delta \hat n^{(1)}(t',x') \rangle  $  which following \cite{bffrc} we approximate as:
\be G_2^{(1)} =\frac{n^{(1)}(x)n^{(1)}(x')}{m^2c^2(x)c^2(x')}  \lim_{t\to t'} \sqrt{\frac{m^2c(x)c(x')}{n^{(1)}(x)n^{(1)}(x')}}
D  \langle  \delta \hat \theta^{(2)}  \delta \hat \theta^{(2)} \rangle
  \label{cinque}\ \ \ \ \ \ \ \ \  \ee
where $D$ is the differential operator $D=(\partial_t+V\partial_x)(\partial_{t'}+V\partial_{x'})$ and $\langle \delta \hat \theta^{(2)}(t,x) \delta \hat \theta^{(2)}(t',x')  \rangle$  is the two-point function of a 2D massless scalar field $\delta\hat\theta^{(2)}$ ($=\sqrt{\frac{n^{(1)}}{mc}}\delta\hat\theta$ 
) propagating in the 1+1D metric
\be ds^2 =  -( c^2 - V^2 ) dt^2  + 2|V|dtdx   +dx^2   \label{sei}\ee
and satisfying
\be \Box  \delta \hat \theta^{(2)}  =0 \ , \label{sette}\ee
where the 2D $\Box$ is calculated from (\ref{sei}).
The approximation used is familiar in QFT in curved space-time when dealing with Hawking BH evaporation in spherically symmetric spacetimes (Schwarzschild for example): the Unruh vacuum (the quantum state that describes Hawking radiation \cite{unruh76}) 4D stress tensor $T_{ab}$ is approximated by $\frac{t_{ab}}{4\pi r^2}$, where $t_{ab}$ corresponds to the stress tensor of a 2D massless scalar field (see for instance \cite{dfu, bb, pp}).
The conformal factor $\frac{n}{mc}$ of the transverse ($y,z$) space plays the role of $r^2$.
The approximation introduced in eq. (\ref{cinque}) makes the model analytically solvable and although one neglects the backscattering of the modes
(studied in \cite{abfp}), the presence and basic features of the main correlation peak can be nicely reproduced.

Since our 2D field is conformally invariant, its vacuum zero temperature two points function has the same form as in Minkowski space-time, namely     $\ln ( \Delta x^+  \Delta x^-  )$ where $x^{\pm}$ are the null coordinates associated
to the spacetime metric \cite{fn}.
We have (up to a, irrelevant in our case,
diverging constant related to the infrared divergence of our 2D theory)
\be \langle \delta \hat \theta^{(2)} (t,x)  \delta \hat \theta^{(2)}(t',x') \rangle  = -\frac{\hbar}{4\pi}  \ln ( u_{in}-u'_{in}) (v_{in}-v'_{in}) \ ,     \label{otto} \ee    
where
\be u_{in} = t - \frac{x}{c_{in}-|V|},\ \   v_{in}= t+ \frac{x}{c_{in}+|V|} \label{nove} \ee                                           
are the null coordinates associated to the metric (\ref{uno}) for $t<0$. This two point function characterizes our ``in''  quantum state for the field  $\delta \hat \theta$ which corresponds to an initial vacuum (i.e. no incoming phonons from both left and right past null infinity).

This expression can be extended for $t>0$ simply by matching the null coordinates along the spacelike shell at $t=0$ (see Fig.(\ref{uno})).  

\begin{figure}
\includegraphics[width=0.7\columnwidth]{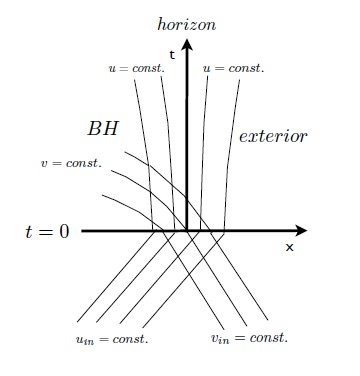}
\caption{Matching of null coordinates between the $t<0$ region, where the flow is uniform and subsonic, and the inhomogeneous $t>0$ region, with a subsonic ($x>0$) and a supersonic ($x<0$) part, representing, respectively, the exterior and the interior of the sonic hole. The horizon is at $x=0$.
 \label{funo}}
\end{figure}

Note that our theory is conformal invariant and so there is no scattering both inside the condensate and at the transition layer.
For $t>0$, the corresponding null coordinates are
\bea u &=&  t-\frac{1}{\kappa}\ln\sinh \frac{3\kappa|x|}{2|V|}, \label{dieci}
\\  v&=& t+\frac{1}{8\kappa}\Big[\frac{9\kappa x}{2|V|}-\ln\cosh\left( \frac{3\kappa x}{2|V|}+\tanh^{-1}\frac{1}{3}\right)\Big]\ . \nonumber     \eea
Matching $u$ and $u_{in}$ at $t=0$ we get
\be -\kappa u =\ln\sinh|Au_{in}| \label{undici} \ , \ee
where $A=\frac{3\kappa}{2|V|}(|V|-c_{in})<0$,
which can be inverted giving
\be |u_{in}A|=\ln( \sqrt{1+e^{-2\kappa u}}+e^{-\kappa u})\ . \label{tredici} \ee                                                                             
For the advanced null coordinates, the choice of the profile also allows to invert the relation $v=v(v_{in})$ and we obtain
\be v_{in}=\frac{4\kappa v}{B}+\frac{1}{2B}\ln\left( 1+\sqrt{1+2\sqrt{2}e^{-8\kappa v}}\ \right)
\ ,     \label{quattordici} \ee
where $B=\frac{ 3\kappa(c_{in}+|V|)}{2|V|}$.
Given these relations, the correlator (\ref{cinque}) for $t>0$ can be analytically evaluated starting from the expression
\bea &&
 G_2^{(1)}(t;x, x')
 = -\frac{\hbar n^{(1)}}{4\pi  mc(x)^{1/2}c(x')^{1/2}}\times \label{quindici} \\
&& \Big[ \frac{1}{(c(x)-|V|)(c(x')-|V|)} \frac{du_{in}}{du}\frac{du'_{in}}{du'}  \frac{1}{(u_{in}-u'_{in})^2} + \nonumber \\ && \frac{1}{(c(x)+|V|)(c(x')+|V|)}  \frac{dv_{in}}{dv} \frac{dv'_{in}}{dv'} \frac{1}{(v_{in}-v'_{in})^2}  \Big] |_{t=t'}  \ . \nonumber \eea
The resulting expression in terms of ($t,x$) coordinates is rather long and will be given elsewhere \cite{p-prep}.
For $t<0$ a similar expression holds, just replace $c(x)$ by $c_{in}$ and ($u,v$) by ($u_{in},v_{in}$).
In Fig. (\ref{due})  we have represented the correlator for points $x>0$ (outside the horizon) and $x'<0$ (inside the horizon) at four increasing times ($t_1=\frac{1}{\kappa}, t_2=\frac{2}{\kappa}, t_3=\frac{4}{\kappa},t_4=\frac{5}{\kappa}$) .

\begin{figure}
\includegraphics[width=0.9\columnwidth]{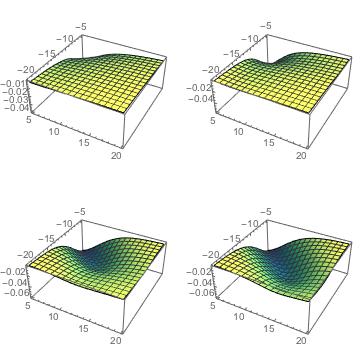}
\caption{Ramp-up of the density correlator (\ref{quindici}) 
 for $5<x<20$, $-20<x'<-5$ and at four different times:
 $t_1=\frac{1}{\kappa}$ (top left), $t_2=\frac{2}{\kappa}$ (top right) and  $t_3=\frac{4}{\kappa}$ (bottom left) and $t_4=\frac{5}{\kappa}$ (bottom right). Here and in all the figures that follow, we have plotted the correlator up to the overall factor $\frac{\hbar n^{(1)}}{4\pi m}$ and chosen
the values $\kappa=\frac{1}{4}, |V|=1, c_{in}=\frac{3}{2}$.  
\label{fdue}}
\end{figure}

One sees very nicely the ramp-up and the formation at late time of the deep valley (the correlator is negative) located at $x=-x'$. This is the signal of Hawking's particle-partner pairs production.
This  can be confirmed analytically by taking the late time limit, $u_{in} \to 0$, $u\to  +\infty$ of eq. (\ref{tredici}) yielding $|u_{in}A| \sim e^{-\kappa u}$ and, from (\ref{quattordici}), $v_{in} \sim  \frac{4\kappa v}{B}$  giving 
\bea
&& G_2^{(1)} \to  \frac{-\hbar n^{(1)}}{4\pi  m |V|^3 \sqrt{(1+\frac{2}{3}\tanh\frac{3\kappa x}{2|V|})(1+\frac{2}{3}\tanh\frac{3\kappa x'}{2|V|})}}\times \nonumber  \\
&& \Big[ -\frac{9\kappa^2}{16 \tanh\frac{3\kappa x}{2|V|}\tanh\frac{3\kappa x'}{2|V|}\cosh^2 \left( \frac{1}{2}\ln \frac{ \sinh \frac{3\kappa x}{2|V|}}{\sinh \frac{3\kappa |x'|}{2|V|}}\right)}+  \label{sedici} \\ && \frac{1}{\sqrt{(2+\frac{2}{3}\tanh\frac{3\kappa x}{2|V|})(2+\frac{2}{3}\tanh\frac{3\kappa x'}{2|V|})}\Big( v(t,x)-v'(t,x') \Big)^2}   \Big]   \ , \nonumber \eea
where in the second term (see the second of (\ref{dieci})) \be v(t,x)-v'(t,x')=\frac{9}{16|V|}(x-x')-\frac{1}{8\kappa}\ln \frac{2e^{\frac{3\kappa x}{2|V|}}+e^{-\frac{3\kappa x}{2|V|}}}{2e^{\frac{3\kappa x'}{2|V|}}+e^{-\frac{3\kappa x'}{2|V|}}}\ .\ee
For $x=-x'$ sufficiently far away from the horizon, the $tanh$ terms  appearing in (\ref{sedici}) can be well approximated by their asymptotic values, while the remaining $\frac{1}{cosh^2}$ term has indeed a minimum for $x=-x'$.
In Fig. (\ref{fquattro}) we plot the value of the correlator (\ref{quindici}) at $t=\frac{10}{\kappa}$ as a function of $x$ for, respectively, fixed $x'= -5,-6,-7,-8$.
\begin{figure}
\includegraphics[width=0.8\columnwidth]{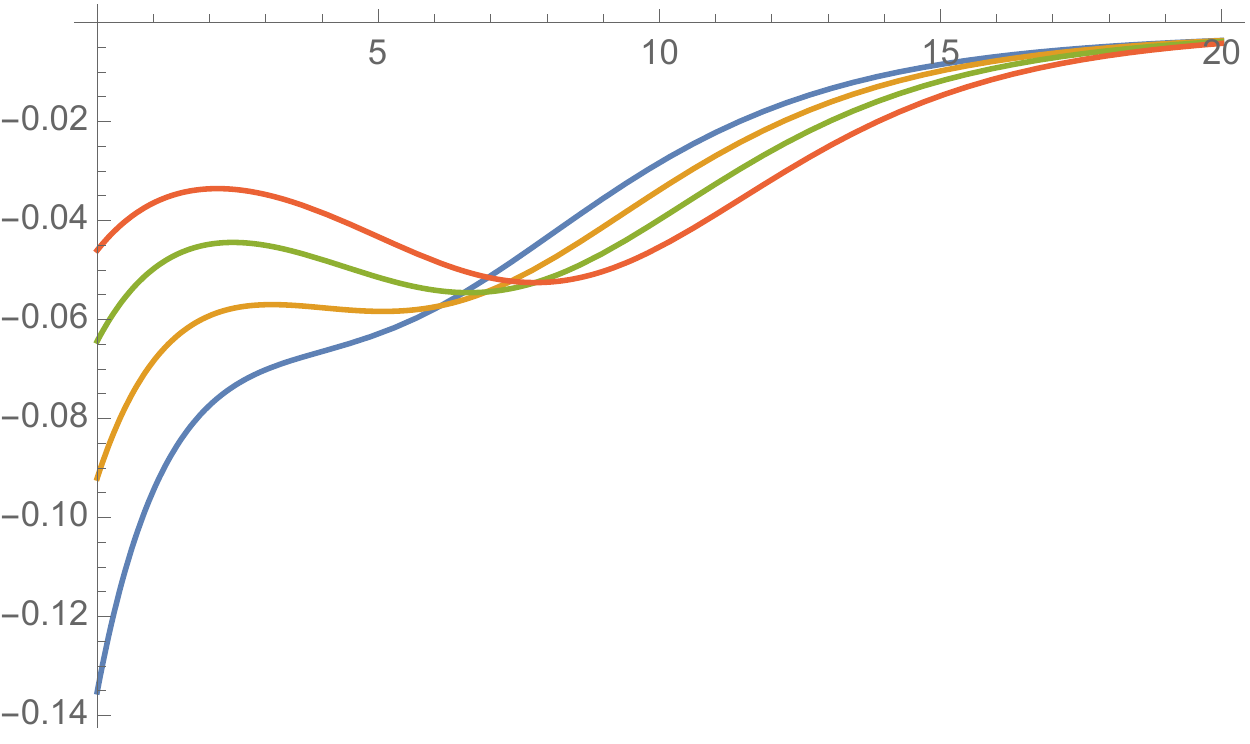}  
\caption{Plot of the density correlator  (\ref{quindici}) 
at $t=\frac{10}{\kappa}$ as a function of $x$ for fixed $x'=-5$ (blue line), $x'=-6$ (orange), $x'=-7$ (green), $x'=-8$ (red).
\label{ftretre}}
\end{figure}
One clearly sees that the peak is located at $x=-x'$ for $|x'|\gtrsim 7$, while for values of $|x'|$
closer to the horizon the peak at $x=-x'$ does not appear. This is due to the fact that as one approaches the horizon ($x=-x' \to 0$) the singularity of the two point function at coincidence points starts dominating \cite{schutzholdunruh}. Thus, pairs production appears not to be located near the horizon but in a region outside it. This is in agreement with previous suggestions on the the existence of  a ``quantum atmosphere'', as referred in \cite{qatm1,qatm2,qatm3}, where Hawking radiation emerges out of vacuum fluctuations.
 In Fig. (\ref{ftran}), obtained by cutting the density correlator (\ref{quindici}) along the line $x=x'+16$, perpendicular to the 
peak at $x=-x'=8$, we plot the temporal formation of the peak profile.
We note that the late-time limit is governed by the condition (one for each point) $e^{-\kappa t}\sinh \frac{3\kappa |x|}{2|V|}=cst\ll1$.
Such $t=t(x)$ governs the formation of the peak at $x=-x'$ and also its length. Away enough from the horizon, the length of the peak grows linearly in $t$, as noticed also in \cite{cfrbf}.
\begin{figure}
\includegraphics[width=0.8\columnwidth]{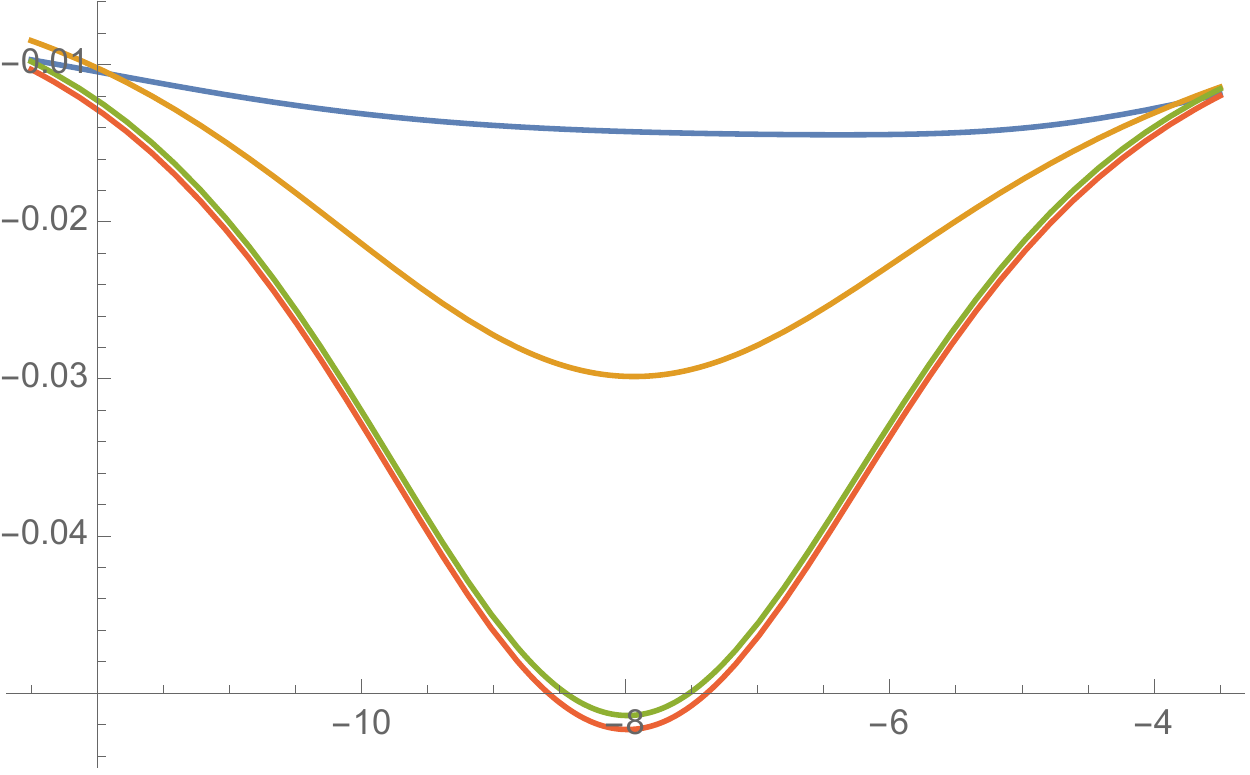}
\caption{Plot of the temporal formation of the peak profile [$t_1=\frac{1}{\kappa}$ (blue), $t_2=\frac{2}{\kappa}$ (orange) and $t_3=\frac{4}{\kappa}$ (green) and $t_4=\frac{5}{\kappa}$ (red)], obtained by cutting  (\ref{quindici}) 
along the line $x=x'+16$, perpendicular to the valley $x=-x'$ at the point $x=-x'=8$.
\label{ftran}}
\end{figure}

Let us now consider the case in which the condensate has an initial temperature $T$, so instead of the vacuum we have, for $t<0$, a thermal distribution of phonons. This analysis is rather important since experimentally a condensate has always a non vanishing temperature which may be comparable or even bigger than the Hawking one $T_H = \frac{\hbar\kappa}{2\pi k_B}$  associated to the thermal emission of phonons by the sonic horizon.
The initial population of phonons in thermal equilibrium in the comoving frame is characterized by an occupation number
\be    N_{\omega_{u(v)}} =   \frac{1}{e^{\frac{\hbar\omega_{u(v)}}{k_BT}}-1}\  \label{diciassette}    \ee
where $\omega_u$ and $\omega_v$ are the Doppler rescaled frequencies corresponding to right moving phonons ($u$)
and left moving ones ($v$)
\be    \omega_u = \frac{\omega c_{in}}{c_{in}-|V|}\ , \  \ \omega_v=\frac{\omega c_{in}}{c_{in}+|V|}\ . \label{diciotto} \ee         

\begin{figure}
\includegraphics[width=0.9\columnwidth]{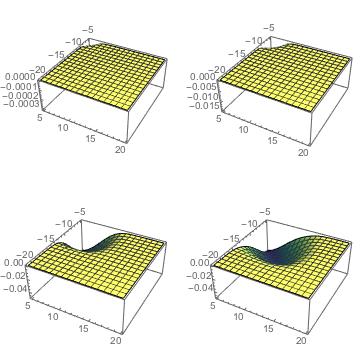}
\caption{Ramp-up of the density correlator  (\ref{venti})
for an initial temperature $T=10T_H$, $5<x<20$, $-20<x'<-5$ and at the four  different times
 $t_1=\frac{1}{\kappa}$ (top left), $t_2=\frac{2}{\kappa}$ (top right) and  $t_3=\frac{4}{\kappa}$ (bottom left) and $t_4=\frac{5}{\kappa}$ (bottom right). 
\label{fsette}}
\end{figure}

The corresponding two-point function for the field  $\delta \hat \theta^{(2)}$ reads \cite{vend}
\be \langle \delta \hat \theta^{(2)} (t,x) \delta \hat \theta^{(2)} (t',x') \rangle  = -\frac{\hbar}{4\pi} \ln \frac{\sinh A_u \Delta u_{in}}{A_u} \frac{\sinh A_v \Delta v_{in}}{A_v},     \label{diciannove} \ee  
where $A_{u(v)}=\frac{\pi k_B T(c_{in}\mp |V|)}{\hbar c_{in}}$,
$\Delta u_{in}=(u_{in}-u_{in}')$ and similarly for $v_{in}$.
The time evolution of the corresponding density-density correlator
\bea
&& G_{2\ T}^{(1)}(t;x,x')
 = -\frac{\hbar n^{(1)}}{4\pi  mc(x)^{1/2}c(x')^{1/2}}\times \label{venti}  \\
&& \Big[ \frac{1}{(c(x)-|V|)(c(x')-|V|)} \frac{du_{in}}{du}\frac{du'_{in}}{du'}  \frac{A_u^2}{\sinh^2 A_u (u_{in}-u'_{in})}   +  \nonumber \\ && \frac{1}{(c(x)+|V|)(c(x')+|V|)}  \frac{dv_{in}}{dv} \frac{dv'_{in}}{dv'}  \frac{A_v^2}{\sinh^2A_v (v_{in}-v'_{in})}   \Big] |_{t=t'}    \nonumber \eea
 is shown
 in Fig. (\ref{fsette}) for
 $T=10T_H$.
 We
 do not see noticeable differences in the time evolution when the BEC temperature
 equals the Hawking temperature ($T=T_H$) with respect to the $T=0$ case.

One sees that the stationary configuration with the valley located at $x=-x'$  ($x\gg 0$) appears even in this case but at a time  
  that
  is bigger than the one required at $T=0$ and this time can increase with $T$.
This can be seen by noticing that at finite temperature the late time limit condition we got at $T=0$ has to be supplemented (for $T>\frac{3c_{in}}{|V|}T_H$)  by the more stringent requirement  $\frac{\pi k_B T(c_{in}-|V|)}{|A| \hbar c_{in}}e^{-\kappa t}\sinh\frac{3\kappa|x|}{2|V|}=cst\ll 1$.
At late times the first term in eq. (\ref{venti}) coming from the $u$ modes contribution reduces to the corresponding one at $T=0$ (the first term in eq. (\ref{sedici})): this is due to the fact that the late time contribution comes only from the modes propagating very close to the horizon (i.e. $u\to + \infty$) and these are highly redshifted, so any information of the initial state is washed out (no hair theorem for Hawking
radiation). Therefore, an initial population causes stimulated emission of $u$ phonons which is however just a transient effect \cite{wald76}. This does not hold for the $v$ modes because they are not redshifted and hence the second term in eq. (\ref{venti}) shows a temperature dependence also in the stationary regime at late time.
The thermal $v$ contribution to the correlator is positive, but, as the $T_H=0$ one (which is negative), smaller than the $u$ contribution. For instance, for $x=-x'=10$ such contributions are one order of magnitude smaller than the $u$ contribution.
So the temperature corrections, within the hydrodynamical approximation (gravitational analogy), 
are small.

In conclusion, in this paper we have presented a simple analytical model able to describe how the Hawking signal in a BEC
emerges  out of a region significantly outside the horizon.
This gives a solid support, now based on the study of correlation functions, on a``quantum atmosphere" as locus of origin of Hawking radiation. 
So far its existence was suggested analyzing the behavior of the expectation values of the stress tensor for quantum fields in a Schwarzschild BH \cite{qatm1,qatm2,qatm3} in the region exterior to the horizon.  However these quantities involve renormalization to cure ultraviolet divergences and include both vacuum polarization and Hawking radiation contributions and it is hard to disentangle them in the region close to the horizon where they are of comparable magnitude.  We think that the formation and subsequent evolution  of the peak in the correlation function that we studied here is a more genuine way to characterize the region where Hawking particles and their corresponding partners materialize out of the vacuum fluctuations.
For a quantitative characterization of this region,
as can be seen in Fig. (\ref{fdue}), the  
  appearance of the signal is not immediate, one has to wait a time of order $\frac{4}{\kappa}$ to see it and it doesn't occur close to the horizon but at a distance of order $\frac{7|V|}{4\kappa}$ (i.e. $x=-x'\sim 7$ in our plots, see the green line in Fig. (\ref{ftretre})) away from it.
As can be seen in Fig. (\ref{sette}), for a BEC ambient temperature $T=10T_H$ 
this delay is enhanced:
the ramp-up process is slower and it lasts up to a time of roughly $\frac{5}{\kappa}$.
We hope this work can be the starting point for a more detailed theoretical modeling based on the powerful methods of QFT in curved spacetime in support of ongoing cold atoms experiments and others involving externally driven or backreacting analog BH horizons.

\textbf{Acknowledgments.}  We thank P. R. Anderson, I. Carusotto, R. Parentani, N. Pavloff and J. Steinhauer for useful discussions.
We also thank anonymous referees for valuable suggestions for improving the Letter. A.F. acknowledges partial financial support by the Spanish Mineco grant FIS2017-84440-C2-1-P and the Generalitat Valenciana grant PROMETEO/2020/079.

\end{document}